# Optimization of Electrolyte Rebalancing in Vanadium Redox Flow Batteries

Mehdi Jafari, *Member, IEEE*, Apurba Sakti, and Audun Botterud, *Member, IEEE*

*Abstract*-- This paper presents a novel algorithm to optimize energy capacity restoration of vanadium redox flow batteries (VRFBs). VRFB technologies can have their lives prolonged through a partially restoration of the lost capacity by electrolyte rebalancing. Our algorithm finds the optimal "number" and "time" of these rebalancing services to minimize the service cost, while maximizing the revenues from energy arbitrage. We show that the linearized form of this problem can be analytically solved, and that the objective function is convex. To solve the complete problem, we develop a two-step mixed integer linear programming (MILP) algorithm, which first finds the bounds for optimal number of services and then optimizes the number, and time of the services. We then present a theoretical analysis and optimization results for a case study of energy arbitrage in New York ISO.

*Index Terms*-- Flow batteries, degradation, cycle aging, capacity restoration, mixed integer linear programming

## I. Introduction

ENERGY storage is one of the promising solutions for large-scale integration of renewable energies in order to decarbonize electricity generation [1]. Among different electrochemical energy storage technologies, Li-ion batteries are the most commercialized, however, emerging solutions such as vanadium redox flow batteries (VRFBs) also have promising performance and lifetime features that can potentially compete with Li-ion batteries in stationary storage applications [2].

RFBs have decoupled power and energy capacities due to independent sizing of the electrolyte tanks and the reactor, which enables longer durations compared to Li-ion batteries [3]. Among RFB technologies, VRFBs have become the most promising and commercially available solution for grid applications. Similar to other electrochemical energy storages, VRFB capacity degrades over time and cycling due to several factors such as electrolyte contamination, electrode corrosion, and ion crossover [4]. The ion crossover represents the diffusion of vanadium ions across the membrane in two half-cells and leads to imbalance of the vanadium species which causes lower efficiency and energy capacity over long-term of cycling [5]. The lost capacity in this process can be partially restored (reversible part) by electrolyte rebalancing through complete or partial remixing of the electrolyte tanks [6].

While VRFB's degradation behavior has been extensively studied from an electrochemical perspective, the same does not apply to operational decision-making in power systems applications, especially when it comes to electrolyte rebalancing [7], [8]. In this paper, we propose a novel algorithm to optimize VRFB operational decision-making and its rebalancing schedule using energy arbitrage as a test case. Our analytical solution shows that the optimal rebalancing scheduling is a convex problem, which we then use to develop a two-step MILP optimization framework to assign the number of and timing of services while maximizing the arbitrage revenue. Results demonstrate that the algorithm can effectively find the optimal operational and rebalancing service schedules.

## II. Theoretical Calculations of VRFB Electrolyte Rebalancing

Considering reversible and irreversible capacity loss components for a VRFB, we optimize restoring the reversible capacity fade by electrolyte rebalancing. For simplicity, we consider the following assumptions: equal arbitrage opportunity for all days, $r$; linear capacity fade with respect to time and cycling; one mandatory rebalancing service at the end of period $D$; $K$ is the cost of each service; $x$ is the number of services inside period $D$; $n$ is the number of days in each service period, and $Q$ is the restorable capacity at the end of period $D$.

Fig. 1 shows the capacity change with and without rebalancing services for a time period of $D$ days. With the fixed daily arbitrage assumption, the daily cycling will be similar and, therefore, the capacity will decline linearly over time. With no capacity restoration service, the capacity will decline from point "$a$" to point "$b$" in Fig. 1. As the energy arbitrage revenue is directly related to the available cycling capacity, the lost revenue opportunity in this period, $R_L^{NS}$, will be the area of the "$abc$" triangle multiplied to the daily revenue potential, $r$:

$$R_L^{NS} = r(\frac{1}{2}QD) \tag{1}$$

However, if we consider $x$ number of services inside $D$, the lost revenue opportunity will be the sum of the smaller dotted triangles such as "$aef$" with area "B". Considering the linear capacity decline assumption and similar cycling in different days, $n_1 = n_2 = \cdots = n_{x+1}$ and the areas of these triangles are equal. The lost revenue opportunity in this scenario, $R_L^S$, will be the sum of all smaller triangles as follows:

$$R_L^S = r\frac{QD}{2(x+1)} \tag{2}$$

Thus, the potential revenue from performing $x + 1$ capacity

---

restoration services, $R_P^S$, will be the difference of these two areas, i.e. the area "A":

$$R_P^S = \frac{1}{2} rQD\left(\frac{x}{x+1}\right) \tag{3}$$

For $x+1$ number of services, the cost will be $K(x+1)$. Therefore, the net potential revenue will be as in (4).

$$R_{NetP}^S = \begin{cases} -K & x = 0 \\ \frac{1}{2} rQD\left(\frac{x}{x+1}\right) - K(x+1) & \forall x \geq 1, x \in N \end{cases} \tag{4}$$

The function in (4) is convex and its maximum value can be calculated by its derivative with respect to $x$ as shown in (5). Fig. 2 shows the net potential revenue with respect to $x$ for different ratios of the service cost $K$ to the lost revenue opportunity in the no-service case ($rQD/2$). The figure indicates that for any $K$ larger than 25% of ($rQD/2$), the net potential revenue will be negative and there is no economic value in performing rebalancing services. However, for lower service costs, there is positive net potential revenue with a maximum value for the optimal number of services ($x_{opt.}$), which depends on the cost/revenue ratio.

$$\frac{dR_{NetP}^S}{dx} = 0 \Rightarrow x = -1 + \sqrt{\frac{rQD}{2K}} \quad \forall x \geq 1, x \in N \tag{5}$$

This analysis with linear assumptions shows that there is only one optimal number of services leading to maximum revenue. Next, we develop a MILP model with realistic assumptions to find this optimal number of services using real price data.

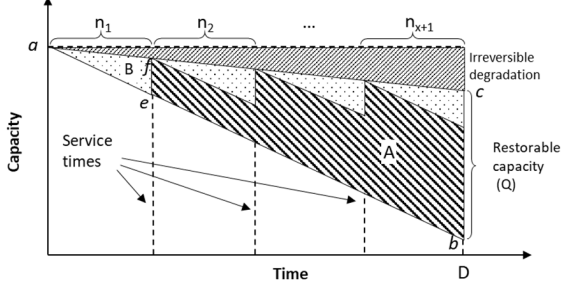

Fig. 1. Linearized capacity evolution over a time period $D$ with reversible and irreversible degradation components

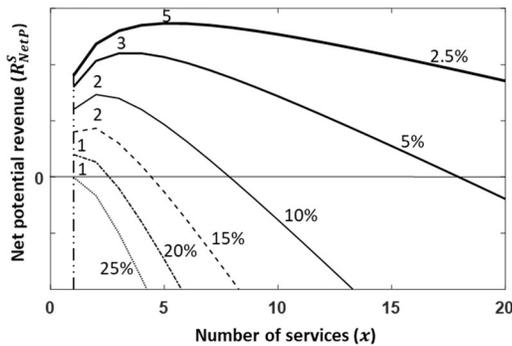

Fig. 2. Net potential revenue vs. number of services for different ratios of the service costs $K$ and the lost revenue opportunity $R_L^{NS}$

## III. MILP Formulation

By varying daily cycling due to price variations, the service periods ($n_1$ to $n_{x+1}$) are not necessarily equal and therefore, the objective is to optimize both the number, and the time of the services. We formulate the objective function as in (6).

$$\max_{E_s, E_p} R = \sum_{y \in Y} \left( \left( \sum_{t \in T_y} \gamma_y(t)[E_s(t) - E_p(t)] \right) / (1+\alpha)^y \right) \tag{6}$$

$$-K(x+1) - C^{VOM} \times \sum_{t \in T}[P_d(t) + P_c(t)]$$

where, revenue is the sold energy $E_s$ minus the purchased energy $E_p$ multiplied by the hourly market price $\gamma$. The net present value (NPV) of the future annual revenues is calculated using the annual discount rate $\alpha$ and summed up to calculate the total revenue. Total cost of services equals $K(x+1)$. Also, the variable operation and maintenance (VOM) cost of the battery is deducted from the total revenue. Counting the number of service days, we have:

$$\sum_{d \in D} b(d) = x + 1 \tag{7}$$

$$b(D) = 1 \tag{8}$$

where, $b$ is a binary variable to define service day. Constraint (8) defines the mandatory service at the end of period $D$. Both calendar and cycle degradations contribute to the total capacity fade and impact the maximum revenue. However, rebalancing services do not affect the irreversible capacity fade. Therefore, the total daily capacity fade, $q$, can be defined as follows:

$$q(d) = 0 \qquad d = 1 \tag{9}$$

$$q(d) = q_{cal}(d) + q_{cyc}(d) \qquad d \geq 2 \tag{10}$$

$$q_{cal}(d) = q_{cal}(d-1) + \frac{\rho \times EOL}{L_{cal}} \qquad d \geq 2 \tag{11}$$

$$q_{cyc}(d) = q_{cyc}(d-1) \times (1 - \sigma b(d)) + (1-\rho) \times EOL \times \frac{\sum_{t \in T_d} P_d(t)}{C_r \times L_{cyc}} \qquad d \geq 2 \tag{12}$$

where, the capacity fade $q$ (p.u) in day $d$ is the sum of calendar $q_{cal}$ and cycle $q_{cyc}$ components in (10). $q_{cal}$ is calculated from the capacity fade in day $d$ based on end of life (EOL) criteria and the calendar life of the battery in days ($L_{cal}$). $\rho$ is the split ratio between calendar and cycle degradations. The method for cycle degradation calculation in (12) is to count the number of equivalent full cycles and compare it to the cycle life ($L_{cyc}$) of the battery. In each day, the discharging power $P_d$ is summed up and divided by the battery's rated capacity ($C_r$). Note that the binary variable $b$ is added to the equation to restore $\sigma$ % of the lost capacity (partial restorations) at the day of service. The total number of cycles at the end of period and the battery capacity in each day are constrained as follows:

$$\sum_{t \in T} P_d(t)/C_r \leq L_{cyc} \tag{13}$$

$$q(d) \leq EOL \qquad \forall d \tag{14}$$

We impose additional constraints to define the battery's cycling in the energy arbitrage problem such as SOC calculations and power constraints, as defined in [9]. In this formulation, $x$ is unbounded and to solve the problem, its upper and lower bounds should be defined to increase the computational efficiency. For a conservative lower bound, it is obvious that $x$ cannot be negative, therefore $x \in W$. The upper bound, however, is not straight forward to define, as it is dependent on the lost revenue due to capacity fade and the service cost. We assume an upper bound number ($X$):

$$x \leq X \qquad \forall x \in W \tag{15}$$

Solving the full optimization problem will optimize both the number and the time of services for capacity restoration. However, to guarantee the optimal solution, $X$ should be defined as a conservatively high number leading to a wide range of possibilities for the binary variable $b$, which is

22

computationally inefficient. Therefore, to reduce the computational burden, we propose a two-step optimization method, such that instead of solving the problem for the whole solution set, we first find the optimal number of services with fixed service periods ($n_1 = n_2 = \cdots = n_{x+1}$), and then, for a reasonable bound on the optimal number of services ($\theta$), we optimize the service number and times. The steps to execute the proposed method are presented in Algorithm 1. With the defined inputs, at step (1), starting from the lowest value for $x$, we assign periodic service times through $b$ and optimize the problem for each $x$ and repeat the process in a do-while loop for ascending values of $x$ until we obtain its optimal value. Then, in step (2), we define the bound of $x$ and solve the problem for optimal scheduling to maximize the total revenue.

---
**Algorithm 1:** Two-step optimization of $x$ and $n_1$ to $n_{x+1}$
---
**Input:** Energy market price data for the time period $t = 1 \ldots T$, the battery performance and degradation parameters, service cost $K$, etc.
Set $x = 0$; $b(i) = 0$: $i = 1 \ldots D$; $R = 0$

(1) do {Set $R_x = R$, $n = \left[\frac{D}{x+1}\right]$
  for $i \in x+1$ do
    $b(i \times n) = 1$
  end
  **Optimize** (6), having $b$ for every $x$
  Save objective value $R$
  Set $x = x + 1$
} while $R > R_x$
Save $x$

(2) Set $X = x - 1$
  **Optimize** (6) subject to all constraints and $X - \theta \le x \le X + \theta$

**Output:** $R$, $x$, $n_1$ to $n_{x+1}$, $E_s$ and $E_p$
---

## IV. ILLUSTRATIVE CASE STUDY

We illustrate the performance of the proposed algorithm with a 0.25MW, 1MWh VRFB using two years of day-ahead market price data (2019-2020) from New York ISO (NYISO). Table I shows the parameters used for optimization [9]–[11]. Fig. 3 shows the total and net revenues for the two-step optimization algorithm with two different service costs. Results of step (1) for 1-10 services and the optimal solution from step (2) are shown. For illustrative purposes, we show results for low services costs of $500 and $200, where the optimal number of services leading to the maximum net revenue is 2 and 3, respectively. For the numbers smaller than the optimal, the arbitrage revenue opportunity is lost due to the faded capacity, and for the number larger than the optimal, the extra revenue does not compensate the added service cost. Note that in the optimal solution, the revenue from two-step optimization (bar graph) is slightly higher than revenue from the step 1 (line graph), due to optimal selection of the service days. Fig. 4 shows the optimal service days in these two cases with reversible and irreversible capacity fades. It is obvious that the days between two services are not necessarily equal due to varying energy prices and different arbitrage opportunities.

Table I. Optimization parameters

| Item | Unit | Value |
| --- | --- | --- |
| Charge/Discharge efficiency | % | 89.7 / 78.6 |
| Calendar life | year | 10 |
| Cycle life | number of cycles | 20,000 |
| EOL | p.u. | 0.3 |
| Discount rate ($\alpha$) | % | 70 |
| VOM | $/MWh | 0.3 |

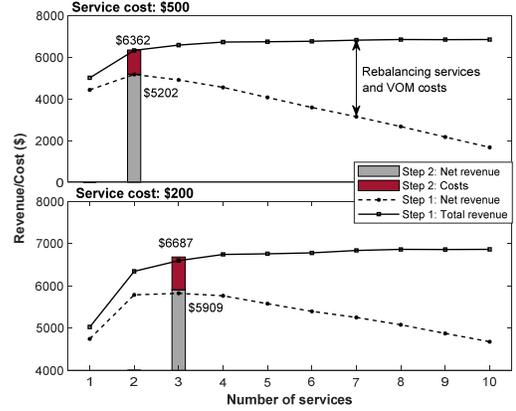

Fig. 3. Revenues and costs in Step 1 and 2 of the proposed algorithm for different service costs ($500 and $200).

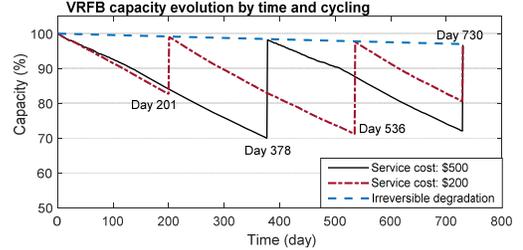

Fig. 4. Optimal VRFB capacity restoration times for different service costs.

## V. CONCLUSION

In this paper, we proposed a two-step optimization algorithm to schedule VRFB electrolyte rebalancing services to maximize the revenues from participating in the energy market. We considered two capacity fade components – reversible and irreversible – and developed a MILP model to define the optimal number and days of rebalancing services. We analytically showed that the simplified linear problem is convex and has one optimal solution. Our results demonstrate that the proposed algorithm can successfully find the optimal service schedule leading to maximum arbitrage revenue.